# A Universal Controller for Grid-Tied Inverters

**Fariba Fateh**, *Senior Member of IEEE*, e-mail: fateh@ksu.edu

**Abstract:** This paper presents the development of "Control-Sync," a novel firmware for universal inverters in microgrids, designed to enhance grid stability and flexibility. As hybrid PV-battery systems become increasingly prevalent, there is a critical need for inverters capable of efficiently transitioning between grid-forming (GFM) and grid-following (GFL) modes. Our firmware introduces dual control paths that allow for seamless transitions without reliance on external control devices, reducing communication overhead and increasing operational reliability. Key features include direct phase-angle detection and frequency restoration capabilities, essential for managing asymmetrical power grids and dynamic load changes. The efficacy of Control-Sync is demonstrated through rigorous testing with grid emulators and multi-phase inverters, confirming its potential to improve microgrid reliability and efficiency. This study offers a scalable solution to enhance inverter adaptability in various grid conditions, fostering a more resilient energy infrastructure.

**Introduction:** According to global estimates, the market for PV inverters is expected to experience a compound annual growth rate (CAGR) of approximately 5.8% from 2022 to 2030. Based on this projection, it is predicted that the market size will reach $17.9 billion by the year 2030 [1]. Also, hybrid PV-battery systems will become a part of residential, commercial, and public buildings along with the rapid expansion of EV market. On the other hand, many experts believe that one essential step toward an energy-efficient world lies in using wide-bandgap (WBG) power semiconductor devices. The technology of WBG devices, such as SiC-MOSFETs and GaN transistors, has significantly enhanced in recent years, with global growth of CAGR ~35.12% from 2021 to 2026 [2]. The adoption of SiC-MOSFETs and GaN transistors in inverters is anticipated to enhance their performance, increase power density, and contribute to overall improvements in the reliability and efficiency of PV systems.

Reliable access to electricity is at risk due to extreme weather events and cyberattacks. In order to ensure that hybrid PV-battery systems can be dependable sources of power, their inverters need to have the ability to function in two modes: grid-following (GFL) and grid-forming (GFM). When operating in GFL mode, the inverters track the grid voltage and generate a PWM reference signal to regulate the active and reactive power that is fed into the utility grid [3], [4]. Inverters in GFM mode, internally generates PWM reference signal, can form a microgrid and regulate the node voltages in cooperation with other power sources to supply the local load. Herein, a microgrid refers to a subdivision or a building that can operate off-grid after disruptive events. The cooperation between power generation units can be achieved by a decentralized or centralized control scheme. If a droop-based decentralized control scheme is implemented, the system frequency can deviate from its nominal value after any load change. In contrast, a centralized supervisory control scheme requires a communication network, which increases the overall cost and risk of cyberattacks and thus compromises system reliability.

A WBG-based universal inverter is an attractive option for hybrid PV-battery systems capable of operating off the utility grid. The universal inverter can seamlessly switch between GFL and GFM modes, stay synced with other power generation units, and contribute collaboratively to the total load in GFM mode. The universal inverter in an islanded microgrid can cooperatively restore the system's frequency and individually restore node voltages without using any communication network.

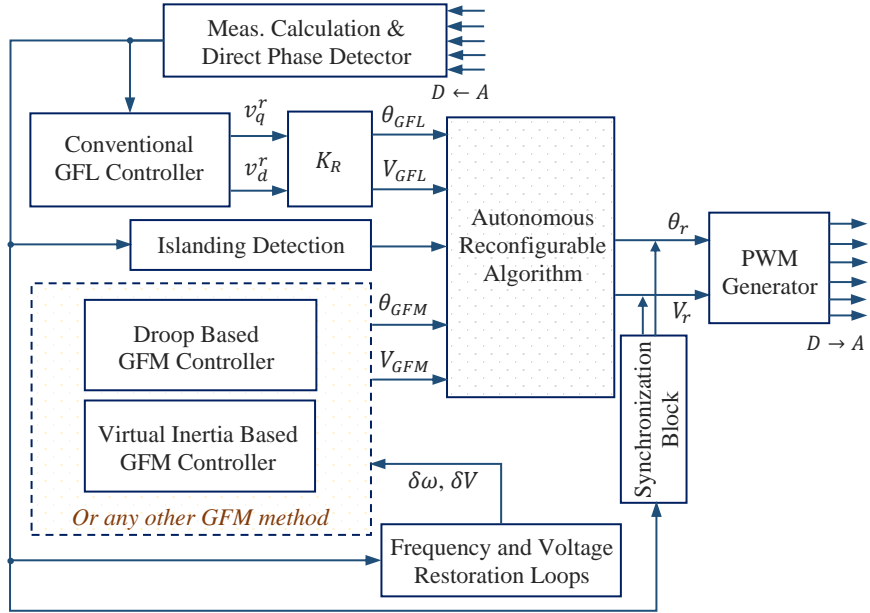

**Figure 1.** Main blocks in the control firmware package for universal inverters.

**Control-Sync. Method:** In this study, we have developed firmware to autonomously and seamlessly generate reference signals for GFM and GFL applications, specifically ($\theta_{GFM}$, $V_{GFM}$) and ($\theta_{GFL}$, $V_{GFL}$). These references are used to implement PWM signals and those are periodically updated, ensuring efficient and accurate performance [5], [6], [7], and [8].

In our research, we successfully developed and implemented a novel control methodology for grid-interactive inverters, which we have named "Control-Sync." This method distinguishes itself by utilizing two separate pairs of control paths to manage the voltage amplitude and phase-angle references. Specifically, one pair is designed for grid-forming (GFM) operations, while the other handles grid-following (GFL) operations. At any point, only one of these pairs actively controls the inverter, while the other remains synchronized but inactive in the background.

A key innovation of the Control-Sync approach is its ability to keep the inactive control paths perfectly synchronized with those in use. This synchronization allows inverters, such as those in a microgrid setup, to switch between GFM and GFL modes seamlessly [9], [10]. This transition occurs without the need for communication with a synchro-check relay, which typically manages the connection of the microgrid to the utility grid. Our approach eliminates the reliance on such devices, facilitating a smoother operational transition in both islanded and grid-connected scenarios.

Moreover, our research introduces minimalistic control loops within the control-sync method, enhancing the system's efficiency. We also developed communication-free schemes for detecting islanding and grid reconnection, thereby enabling the inverters to operate autonomously within a microgrid environment.

The effectiveness of the Control-Sync method was rigorously tested and validated using a sophisticated hardware setup that included two 30-kVA grid emulators and four 5-kVA, 208-V three-phase inverters. This practical application not only demonstrated the feasibility of our method but also its robustness [5]-[8].

Our team conducted a thorough evaluation of potential hardware platforms, using products from Texas Instruments, i.e., TMS320F28379D, which is a 32-bit floating-point microcontroller designed specifically for advanced closed-loop control applications. This microcontroller is particularly suited for tasks such as industrial motor drives and solar inverters, where precise and robust control is paramount.

The team has implemented sections of the control algorithm, as illustrated in the block diagram shown in Figure 1. These implementations are critical in demonstrating the practical applicability of the theoretical constructs developed in this research.

**Features of Universal Controller:** A "weak grid" refers to an electrical grid with limited capacity and stability characteristics, typically due to high impedance or low short-circuit capacity relative to the load or power generation within the grid. In simpler terms, a weak grid is not robust enough to handle large disturbances or fluctuations in power demand and supply, often resulting in voltage fluctuations and instability [11], [12], and [13].

Microgrids are smaller-scale electricity networks that can operate independently or in conjunction with the larger utility grid. They often incorporate renewable energy sources, like solar panels and wind turbines, along with conventional generators and battery storage. The term "weak grid" is applicable to many microgrids because:

1. *Low Inertia:* Unlike large power systems with substantial mechanical inertia from big synchronous generators, microgrids typically rely more on power electronics and smaller, often renewable, generators that provide little to no inertia. This lack of inertia makes the microgrid more susceptible to frequency and voltage fluctuations.
2. *Limited Short-Circuit Capacity:* Microgrids usually have a lower capacity for providing short-circuit current due to the smaller size and power output of their generating units. This limitation makes it more difficult to maintain stable operations during faults or abrupt load changes.
3. *High Impedance:* The smaller scale and lower capacity of transmission and distribution components in microgrids can lead to higher impedances, further contributing to the instability and voltage issues.

The team has examined the stability challenges voltage source inverters (VSIs) face in weak grids, emphasizing the critical need for robust control mechanisms to counteract the variations in grid impedance that significantly affect the inverters' ability to provide stable power [11], [12], and [13]. Building on this, Adib and Mirafzal (2019) introduce the concept of virtual inductance, detailing how it can enhance the performance of grid-interactive VSIs, allowing them to effectively manage swift changes in load and generation, thereby maintaining voltage stability and mitigating oscillation risks in low-inertia power systems [14]. Further, the team has also explored the stability of islanded microgrids, exemplifying low-inertia systems prone to weak grid behaviors, and highlight the necessity of advanced control strategies to maintain stability in the absence of traditional grid support, ensuring resilient operation under both normal and fault conditions [15].

Together, these works collectively emphasize the necessity of sophisticated control strategies in managing the stability and reliability of inverters within weak grid environments, such as microgrids. They provide a strong foundation for further research and development in advanced inverter technologies that can adapt to and overcome the limitations of weak grid infrastructures.

These findings are crucial for designing future power systems that are both resilient and efficient, capable of integrating a higher proportion of renewable energy sources while maintaining grid stability. These features have been integrated in the universal inverter controller.

Ancillary services are essential for universal inverters in microgrids to ensure stability, efficiency, and reliability in power generation and distribution. These services include voltage and frequency regulation, reactive power support, and the ability to maintain system integrity under dynamic conditions. The necessity for these services arises from several challenges and operational demands highlighted in key studies:

1. *PWM Common Mode Reference Generation* (Lamb, Mirafzal, and Blaabjerg, 2018) discusses maximizing the linear modulation region of CHB (Cascaded H-Bridge) converters in islanded microgrids. This optimization is crucial as it enhances the quality and stability of the output power, which is particularly significant in the absence of a strong grid's regulatory influence [16].
2. *Ancillary Services via VSIs in Microgrids* (Adib, Lamb, and Mirafzal, 2019) elaborates on how Voltage Source Inverters (VSIs) can provide ancillary services to microgrids, including maximum DC-bus voltage utilization. This capability is vital for maintaining voltage levels within specified limits, thereby ensuring the operational stability of the grid and the optimal functioning of connected devices [17].
3. *Grid-Interactive Cascaded H-Bridge Multilevel Converter PQ Plane Operating Region Analysis* (Lamb and Mirafzal, 2017) focuses on the power quality (PQ) plane operating regions of converters. This study is integral to understanding how inverters can adjust to fluctuating power demands and maintain power quality, which is essential for preventing disruptions in microgrids [18].
4. *An Adaptive SPWM Technique for Cascaded Multilevel Converters* (Lamb and Mirafzal, 2016) investigates a technique for improving the efficiency of multilevel converters with time-variant DC sources, which is crucial for adapting to varying power generation conditions typically seen in renewable energy applications. Such adaptability supports continuous and efficient power supply despite source variability [19].

These ancillary services facilitated by advanced inverter technologies are critical for microgrids, especially when isolated from the main grid. They not only support the basic functionality of providing power but also enhance the resilience and adaptability of the grid to different operational conditions, thereby safeguarding against power quality issues and outages.

Current Source Inverters (CSIs) are particularly adept at handling grid disturbances, making them an excellent choice for grid-tied applications that require robust low voltage ride-through (LVRT) and fault ride-through capabilities. Unlike VSIs, which may need additional complex hardware to cope with voltage dips and faults, CSIs naturally possess inherent fault tolerance [20], and [21]. This is due to their current-driven nature and the large inductors they typically incorporate, which limit the rate of change in current and provide built-in protection against short circuits. This allows CSIs to continue operating smoothly during under-voltage events, maintaining output current independently of the voltage conditions. Additionally, their simplified control systems facilitate easier management of grid disturbances. With enhanced LVRT capabilities, CSIs can support the grid during critical conditions by injecting or absorbing reactive power as needed, thus contributing significantly to the stability and reliability of the power network.

Advanced droop control and smart loads significantly enhance the performance of inverters in microgrids, addressing stability and efficiency in systems with mixed inertia and dynamic power flow. Advanced droop control autonomously adjusts inverter output, aligning frequency and voltage with fluctuations in load and generation, which is crucial for maintaining supply-demand balance, particularly in isolated or disturbed grids [22], [23], and [24]. For example, battery-fed smart inverters utilizing this technology support critical frequency and voltage levels in microgrids, improving resilience and enabling smoother integration of renewable resources. Additionally, smart loads dynamically manage power consumption based on real-time grid conditions, smoothing demand profiles and preventing overloads, which stabilizes the grid and reduces outage risks. Together, these technologies foster a robust, flexible network that adapts to varying energy scenarios, optimizing power distribution and enhancing the overall functionality of microgrids.

Universal inverters in microgrids require several advanced features to ensure their effectiveness and reliability across a variety of grid conditions [25], including asymmetrical and unstable environments. Key features identified in recent research include:

1. *Direct Phase-Angle Detection:* As described by Sadeque, Benzaquen, Adib, and Mirafzal (2021), this feature is crucial for inverters operating in asymmetrical power grids. It allows the inverters to accurately detect and adjust to phase angle variations, which is essential for maintaining stable operation and synchronizing with the main grid despite irregular power flows or grid faults [26].

2. *Frequency Restoration Capabilities:* Highlighted in the work by Sadeque and Mirafzal (2023), frequency restoration is vital for grid-forming inverters, particularly during pulse load and plug-in events. This capability enables inverters to quickly respond to sudden changes in load (such as those caused by large appliances turning on or electric vehicles charging) and stabilize the grid frequency, thereby ensuring consistent power quality and preventing disruptions [27].

3. *Cooperation in Black-Start Conditions:* Multiple grid-forming inverters must effectively collaborate in black-start scenarios—when the grid needs to be restarted after a complete shutdown. The challenges and strategies for such coordination are explored by Sadeque, Sharma, and Mirafzal (2021). This feature ensures that inverters can work together to re-energize and stabilize the grid without external power sources, which is critical for enhancing grid resilience and recovery capabilities [28].

These features highlight the importance of adaptability, quick response, and collaborative operation in universal inverters within microgrids.

Security, particularly device-level security, is an increasingly critical feature for universal grid-tied inverters due to the heightened risks of cyber-attacks and physical tampering in an interconnected power grid environment. These inverters, integral to the operation of electrical grids, require robust protection to ensure reliability, prevent service disruptions, and safeguard against external threats.

1. *Analytical Risk Factor Integration:* As explored by Hossen, Amariucai, and Mirafzal (2024), integrating an analytical risk factor into the operational framework of inverters can enhance their self-protection capabilities. By embedding this risk assessment into a neural network framework, inverters can actively evaluate threats in real-time and adapt their operational strategies to mitigate potential risks, thereby preventing manipulations that could lead to grid instability or failure [29].

2. *Protection Against Malicious Setpoints:* The study by Hossen, Gursoy, and Mirafzal (2022) discusses the implementation of self-protective mechanisms in inverters to shield against malicious control setpoints. Using analytical reference models, these inverters can detect and reject unauthorized or harmful instructions before they affect the grid. This capability is crucial for preventing cyber-attacks that attempt to destabilize the grid by altering inverter behavior, ensuring continuous and secure operation [30].

These features underscore the necessity for advanced security measures in grid-tied inverters. As the grid becomes more digitized and connected, the potential impact of cyber threats grows, making it imperative for inverters to have built-in security features that protect against both cyber and physical vulnerabilities. Such capabilities not only maintain the integrity and functionality of the power grid but also build trust in the infrastructure's resilience against emerging security challenges.